\documentstyle[11pt,newpasp,twoside,epsf]{article}
\begin{document}
\title{Ultraviolet and Visible Spectropolarimetric Variability in P Cygni}
\author{K.  H.  Nordsieck, J. Wisniewski, B. L. Babler, M. R. Meade,
C. M. Anderson, K. S. Bjorkman, A. D. Code, G. K. Fox, J. J. Johnson, A. J. Weitenbeck, N. E.
B. Zellner}
\affil{Space Astronomy Lab, University of Wisconsin, Madison WI 53706}
\author{O. L. Lupie}
\affil{Space Telescope Science Institute, Baltimore MD 21218}

\begin{abstract}
We report on new data, including four vacuum ultraviolet spectropolarimetric observations by
the Wisconsin Ultraviolet Photo-Polarimeter Experiment ("WUPPE") on the Astro-1 and Astro-2 shuttle 
missions, and 15 new visible-wavelength
observations obtained by the HPOL CCD spectropolarimeter at the Pine Bluff Observatory of the
University of Wisconsin.  This includes three HPOL observations made within 12 hours of each
of the three Astro-2 WUPPE observations, giving essentially simultaneous observations
extending from $1500$ to $10500 \AA$.  An analysis of these data yields estimates on the
properties of wind "clumps" when they are detectable by the polarization of their scattered light. 
We find that the clumps must be detected near the base of the wind, $ r/R_{*}  = 1.3 - 2.5 $.  At
this time the clump density is at least $10^{13} \, \mbox{cm}^{-3} $, 20 times the mean wind, and the
temperature is roughly 10,000 \deg K, about $20\%$ cooler than the mean wind.  The clumps
leading to the largest observed polarization must have electron optical depth of 0.1 - 1 and an
angular extent of 0.1 - 1 ster, and account for at most $2\%$ of the wind mass loss. The H and HeI
emission lines from the wind are unpolarized, but their P Cygni absorption is enhanced by a
factor of four in the scattered light. We speculate on the possible relationship of the clumps
detected polarimetrically to those seen as Discrete Absorption Components (DACs) and in
interferometry.
\end{abstract}

\section{Introduction}

P Cygni was found by many early polarimetric investigators to possess intrinsic linear
polarization (eg, Coyne \& Gehrels 1967; Coyne, Gehrels, \& Serkowski 1974; Serkowski,
Mathewson, \& Ford 1975).  Hayes (1985) and Lupie \& Nordsieck (1987) established that the
polarization was variable on timescales of days - weeks, with an amplitude of approximately
0.4\%, and with no apparent periodicity nor favored position angle, although the interpretation
was hampered by an appreciable, unknown interstellar polarization.  The long time series of
Hayes (1985) exhibits a polarimetric episode length of $\sim 1$ week, and a repetition rate of
once per 20-30 days.  Taylor, Nordsieck et al (1991, hereafter "TN"), in the first detailed
spectropolarimetric investigation, made use of the apparent non-variability of the polarization of
the $H\alpha$ emission line to estimate the interstellar polarization there, extrapolating to other
wavelengths using a galactic mean interstellar wavelength dependence.  After removing this
interstellar polarization, TN then confirmed the remarkable daily variability and the random
nature of the degree and position angle of the intrinsic continuum polarization.  The $H\alpha$,
$H\beta$, and $HeI \, \lambda 5876$ lines were found to be essentially unpolarized, while the
continuum polarization wavelength dependence is not flat, but exhibits a steady decrease from
the near ultraviolet into the red. TN pointed out that this, coupled with the lack of a polarization
Balmer Jump, was difficult to explain using the model which is conventionally used for Be stars,
where hydrogen bound-free opacity competes with wavelength-independent electron scattering to
impart a polarization wavelength dependence.

\section{New Spectropolarimetry}

The new polarimetric data presented here includes four vacuum ultraviolet spectropolarimetric
observations by WUPPE on the shuttle missions
Astro-1 (one observation: 12 Dec, 1990) and Astro-2 (three observations: 3, 8, and 12 March,
1995), and 15 new visible-wavelength observations obtained by the "HPOL"
CCD spectropolarimeter at the Pine Bluff Observatory of the University of Wisconsin.  The
WUPPE instrument is described in Nordsieck et al (1994) and HPOL is described in Nordsieck
\& Harris (1995).  The visible wavelength spectropolarimetry  includes three HPOL observations
made within 12 hours of each of the three Astro-2 WUPPE observations, giving essentially
simultaneous observations extending from $1500$ to $10500 \AA\ $.  The Astro-1 WUPPE
observation was discussed by Taylor, Code, et al (1991), who noted a total continuum
polarization generally consistent with interstellar, plus possible line-blanketing of the intrinsic
polarization, an effect similar to that observed in WUPPE observations of Be stars (Bjorkman, et
al 1991).

\begin{figure}
\plotfiddle{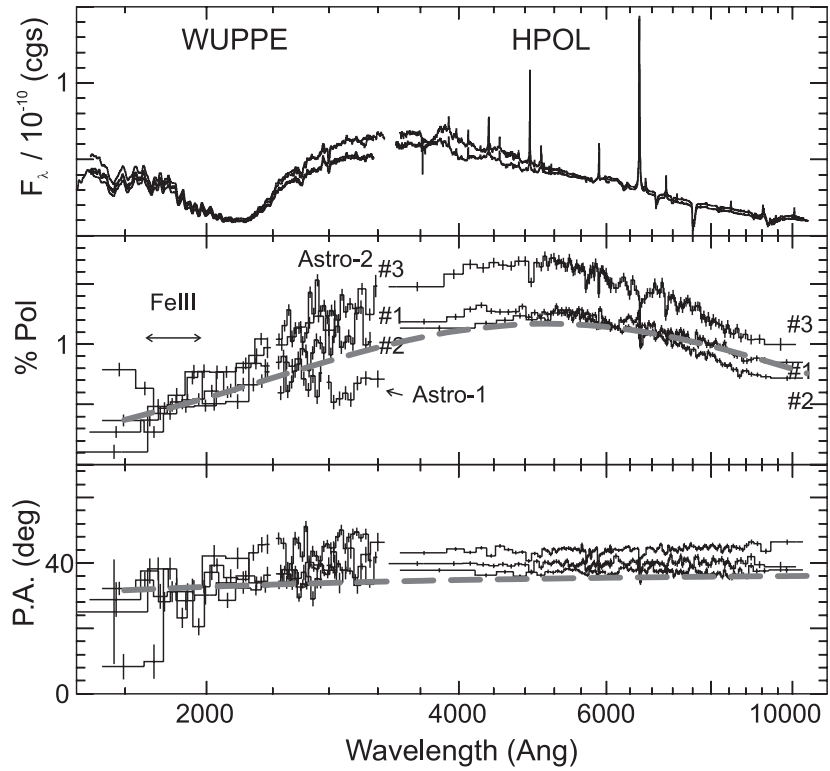}{3.5in}{0}{55}{55}{-165}{-95}
\caption{Observed polarization of P Cyg.  Top: flux.  Middle: degree of linear polarization, for
the Astro-1 and the three Astro-2 observations, with corresponding HPOL observations.  Bottom:
polarization position angle.  Gray dashed line: interstellar polarization estimate}
\end{figure}

Figure 1 shows the net polarization vs wavelength from the WUPPE observations and the three
contemporaneous HPOL observations.  The observations exhibit an overall interstellar
polarization continuum shape plus strong intrinsic polarization variability over most of the
spectrum.  The interstellar polarization is estimated in a variety of ways.  First,  we see that the
four WUPPE observations all cross in the $1700-1900 \AA\ $ region.  We interpret this to be
strong FeIII line blanketing which is suppressing the intrinsic polarization there.  This region is
known to be the most strongly blanketed region in P Cyg (Pauldrach and Puls 1990), and strong
polarization suppression there is seen similarly in the Be star $\zeta$ Tau (Bjorkman, et al 1991).
Conveniently, this then establishes the interstellar polarization value at $1800 \AA$, $0.47 \pm
0.03\%$ at PA 32.1.  Second, following TN, the $H\alpha$ line is assumed to be not intrinsically
polarized.  
\begin{figure}
\plotfiddle{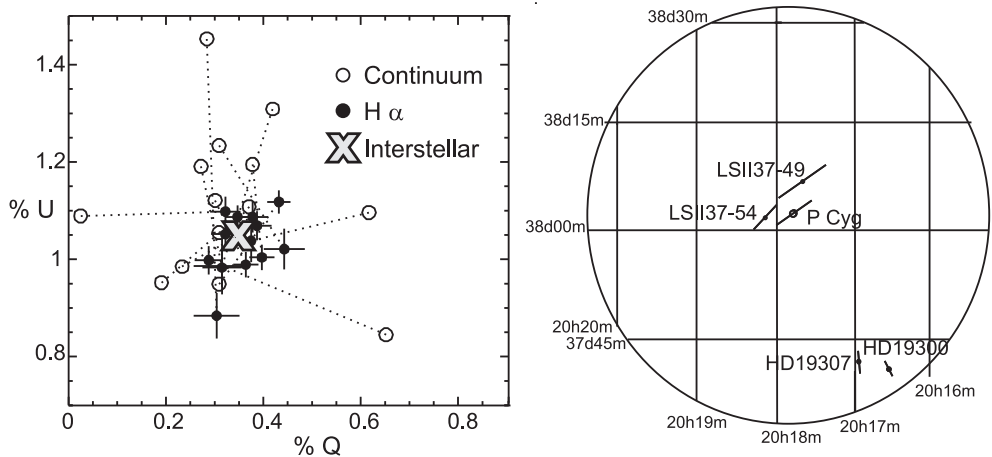}{1.75in}{0}{70}{70}{-185}{-325}
\caption{2a (left): Polarization vector diagram for 15 HPOL observations.  2b(right): Interstellar
polarization map}
\end{figure}
This is supported by figure 2a, which shows a polarization vector diagram for all of the HPOL
observations.  The continuum polarization near $H\alpha$ varies by roughly $\pm 0.3\%$ with
no apparent favored position angle, while the $H\alpha$ points are consistent with a constant
polarization of $1.11 \pm 0.03\%$ at PA 35.5. This is then taken to be the interstellar
polarization at $H\alpha$ (cross in figure 2a).  Using the standard Serkowski interstellar
polarization law as modified by Wilking, Lebowski, \& Rieke (1987; "WLR"), plus a modest
position angle rotation, these two points determine the actual P Cyg WLR parameters: $
p_{max} = 1.17 \pm 0.03\%$, $\lambda_{max} = 5100 \pm 100\AA $, $P.A. = 35.1\deg$,  and a
position angle rotation of $ -0.8 \pm 0.5 \deg/\micron^{-1}$.  The adopted interstellar
polarization curve is shown as the gray dashed line in figure 1.  We have attempted to verify the
above estimate through observations of stars near P Cyg (figure 2b).  Observations with HPOL
with the 0.9m telescope at PBO of HD19300 and HD19307 (17 arcmin distant) exhibit rather
different interstellar polarizations, consistent with the large known variations in Cygnus.  New
observations at the WIYN 3.5m telescope of two B stars less than 5 arcmin distant within the P
Cyg cluster, LSII 37-49 and -54,  yield more comparable results, 1.5\% at $35\deg$ and 0.9\% at
$47\deg$.  This provides support for our adopted value.

\section{Intrinsic Polarization }

\begin{figure}
\plotfiddle{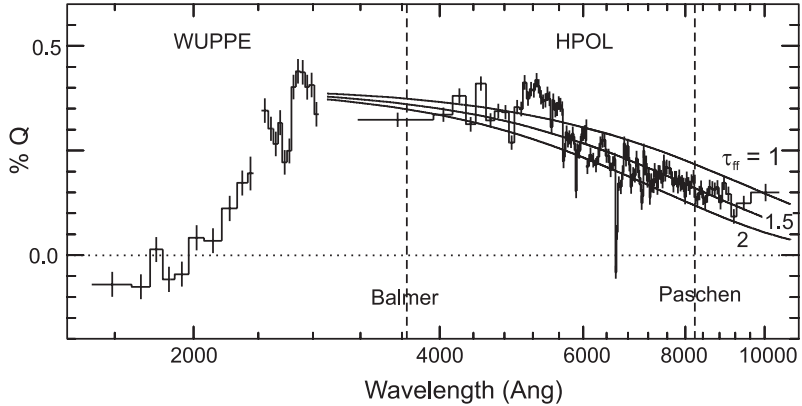}{2.1in}{0}{67}{67}{-235}{-130}
\caption{P Cyg mean intrinsic polarization wavelength dependence.  Fit with competing free-free
absorption for clump optical depth  $ \tau_{\mathit{ff}} (1 \micron) = 1$, $1.5$, and  $2$.}
\end{figure}

In Figure 3 we show the mean intrinsic polarization wavelength dependence for the three Astro-2
+ HPOL observations.  This was obtained by removing the fitted interstellar polarization, rotating
the intrinsic polarization into the Q stokes parameter using the blue position angle, and averaging
the result.  This is the same procedure used by TN to estimate the mean wavelength dependence
of the intrinsic polarization: it assumes that the intrinsic polarization at one time has a position
angle independent of wavelength.  The Stokes polarization  Q  is positive when the polarization
is parallel to the blue continuum polarization, and negative when it is orthogonal. Note the strong
decrease into the vacuum ultraviolet (presumably due to line blanketing), the strong decrease into
the red as noted by TN but here extended into the near infrared, and the lack of both a Balmer
and Paschen jump .

\subsection{Clump Optical Depth}

It is assumed that for a star as hot as P Cyg, any continuum intrinsic polarization is due to
electron scattering in the envelope, presumably in asymmetric inhomogeneities ("clumps"). 
We may estimate the electron scattering optical depth  $ \tau_{e} $  of the clumps from a simple
single-scattering analysis of the degree of polarization.  Using formulae in Cassinelli, Nordsieck,
and Murison (1987), the net polarization from a sector- shaped density enhancement subtending
$\Delta \Omega < 1$ ster and with electron scattering optical depth $ \tau_{e} < 1 $ is
\begin{eqnarray}
p \approx  (4\pi)^{-1} / (1 + (4\pi)^{-1})  \sin^{2} i \, D(r)\, \tau_{e}\, \Delta\Omega \nonumber\\
= 0.073 \sin^{2} i  \, (1 - (R_{\mathit{phot}}/r)^{2} )^{1/2}  \tau_{e}\, \Delta\Omega. \nonumber
\end{eqnarray}
Here $ i $ is the inclination of the clump direction to the line of sight, and $ D(r) $ is the "finite
disk correction factor" which corrects for cancellation of polarization near the stellar photosphere
caused by illumination from a non-point source (here we take the approximation of no
photospheric limb darkening).  Given the rapid timescale and the large amount of polarization
observed, we expect that the clumps are polarimetrically observable only near the base of the
wind. Later we find that the appropriate value for the radial position of the clumps is $ r/R_{*} =
1.4 - 2.5$, which gives $ D \sim 0.4 - 0.9 $ for $ R_{\mathit{phot}}/R_{*} = 1.3 $ (Najarro 2000); we
will take $ D = 0.65$.  This formula holds approximately until $ \tau_{e} \sim 1$, where the
clump becomes optically thick to the total opacity, and $ \Delta\Omega \sim 1$ ster, where
geometric cancellation over the clump begins to reduce the polarization.  Thus the maximum
possible polarization (for $ i = 90$) is  
\begin{displaymath}
p_{max} \sim 0.073 \times 0.65 \, \tau_{e} \, \Delta\Omega \sim 0.05 \, \tau_{e} \, \Delta\Omega.
\end{displaymath}
The observed polarization amplitude below $5000 \AA$ is roughly 0.004, so that  $\tau_e \,
\Delta\Omega \sim 0.1$.  A range of clump properties consistent with the observations is then $
(\tau_{e}, \Delta\Omega) = (0.1, 1)$ to $(1, 0.1)$.

\subsection{Free-Free Absorption - Clump Density}

The presence of continuum absorption optical depth $ \tau_a
$  in addition to the electron scattering optical depth $ \tau_e $ reduces the net polarization by the
absorption in the clump
\begin{displaymath}
p(\lambda) = F_p (\lambda) / F(\lambda) \, \propto \, \tau_e \, e^{ -\tau_a }
\end{displaymath}
where $ F $ is the total flux and $ F_p $ is the polarized scattered flux.  Features in $ p(\lambda)
$ may then be used to estimate  $\tau_{a}(\lambda)$, and thus the physical properties of the
clumps.

An important feature of the observed continuum intrinsic polarization wavelength dependence is
the persistent decrease into the infrared.  This was also noted by TN, and holds for individual
nights as well as the mean.  A plausible explanation for this, given the high density and
ionization at the base of the wind, is competing free-free absorptive opacity.  Since $ \kappa_{\mathit{ff}}
\propto \lambda^2$, the polarization will decrease into the red as $ p(\lambda) \propto e^ {-
a\lambda^2}$.  Figure 3 shows a fit to the polarization for  $ \tau_{\mathit{ff}}(1 \micron) = 1$, $1.5$,
and $2$.  We adopt $1.5$ as the best value.  Given our the range of allowed $ \tau_e = 0.1 - 1$,
we have $ \tau_{\mathit{ff}}(1 \micron)/ \tau_e = 1.5 - 15$. This provides an estimate of the clump
electron density:
\begin{eqnarray}
\tau_{\mathit{ff}}(\lambda) / \tau_e  =  \kappa_{\mathit{ff}} (\lambda) / \sigma 
= \alpha_{\mathit{ff}} \lambda^2 g N_e N_p T^{-3/2} / N_{e} \sigma_T  \quad\mbox{(cgs units),} \nonumber\\
\mbox{or}\quad
N_{e}(\mathit{clump} ) = 3 \times 10^{12} \, T_4 (\mathit{clump})^{3/2}  \tau_{\mathit{ff}}(1 \micron) / \tau_e  \mbox{
cm}^{-3}  \sim 10^{13} \mbox{ cm}^{-3}. \nonumber
\end{eqnarray}
(For these rough estimates we assume a highly ionized pure hydrogen gas, $ T_4 $ is the clump
temperature in units of $ 10^4  \deg K$, and $ g \approx 1 $ is the Gaunt factor).  This is
approximately 20 times the density in the wind at these radii in the model of Najarro (2000).

\subsection{Bound-Free Jumps: Clump Temperature and Ionization}

The lack of Balmer or Paschen Jumps in the intrinsic polarization of figure 3 puts an upper limit 
$ \tau_{\mathit{bf}} (\lambda_B = 3646) < 0.1 $ for neutral hydrogen bound-free absorption in the
clumps, which then means  $ \tau_{\mathit{bf}}( \lambda_B ) / \tau_{\mathit{ff}} (1 \micron) < 1/15$. This
ultimately puts a limit on the clump temperature:
\begin{displaymath}
\frac{\tau_{\mathit{bf}} ( \lambda_B )} { \tau_{\mathit{ff}} (1 \micron)} = \frac{\alpha_{\mathit{bf}} ( \lambda_B ) N_H
(n=2) }{ \alpha_{\mathit{ff}} (1 \micron) g N_e N_p T^{-3/2}} 
= \frac{\alpha_{\mathit{bf}} ( \lambda_B )} { \alpha_{\mathit{ff}} (1 \micron)} \, X_H \, b_2 \, \frac{e^ {-10.2/kT} }{
N_e T^{-3/2} }, 
\end{displaymath}
where $ X_H $ is the neutral fraction and $ b_2  $ is the departure coefficient for the second state
of hydrogen in the clump.  For simplicity we will assume that $ b_2 $ is unchanged in the clump;
it is less than two at the base of the wind in the models of Najarro (2000), in any case.  For $
X_H $, we assume pure optically thin photoionization, so that it may be scaled from the base
wind density and temperature:
\begin{displaymath}
X_H (\mathit{clump}) = X_H (\mathit{wind}) \frac{N(\mathit{clump})}{N(\mathit{wind})} \left(\frac{T(\mathit{clump})}{T(\mathit{wind})}\right)^{-0.8}.
\end{displaymath}
Finally, then,
\begin{eqnarray}
\frac{\tau_{\mathit{bf}} ( \lambda_B ) }{ \tau_{\mathit{ff}} (1 \micron)} = 7.5 \times 10^{19} \,\frac{b_2 \, X_H
(\mathit{wind}) \, T_4 (\mathit{wind})^{0.8}}{ N(\mathit{wind})} \times \nonumber \\
T_4 (\mathit{clump})^{0.7} e^{-11.8/T_4 (\mathit{clump})}  < 1/15 \nonumber
\end{eqnarray}
\begin{figure}
\plotfiddle{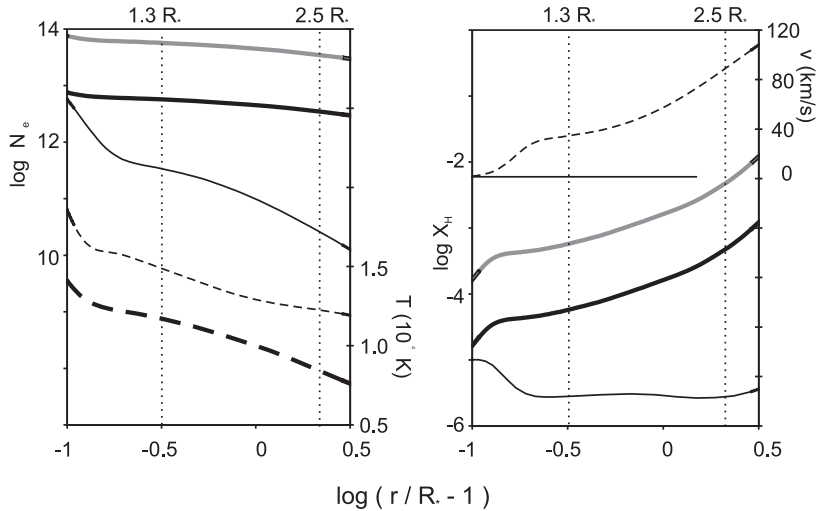}{2.4in}{0}{65}{65}{-195}{-295}
\caption{Possible polarimetric clump properties as a function of radial position.  Left: density
(solid lines, lefthand scale);  temperature (dashed lines, righthand scale).  Right: neutral fraction
(solid, lefthand scale); wind velocity (dashed line, righthand scale). For both, wind: light lines;
clumps: heavy gray lines ($ \tau_e = 0.1$), heavy black lines ($ \tau_e = 1$).  The clump
temperature is the maximum allowed for all clump optical depths.}
\end{figure}
Given a wind model, this may be solved for $T(\mathit{clump})$, giving the maximum temperature of the
clump consistent with no detectable bound-free absorption. Notice that this is independent of
clump electron scattering optical depth and density. Figure 4 (heavy lines) shows the clump
density and ionization corresponding to the maximum temperature at each radius in the wind,
using the wind model of Najarro (2000), for $ \tau_e = 0.1$ and $1$.  The clumps will not be
detectable inside $ r = 1.3 R_* $, the position of the photosphere.  Above $ r = 2.5 R_*$, the
maximum temperature decreases below $ 8,000 \deg K$ and recombination must occur; this puts
a rough upper limit on the radius.  

{\it The clumps must contribute their maximum polarization from $ r = 1.3$ to $ 2.5 R_* $, and
at this time they must be at least 20 times more dense and 20\% cooler than the wind. }

\subsection{Clump Geometry and Time Evolution}

It is interesting to compare the geometric properties of the range of possible clumps. "Thin"
clumps have $ \tau_e = 0.1 $ and $ \Delta \Omega = 1$. "Thick" ones have $ \tau_e = 1 $ and $
\Delta \Omega = 0.1$.   The physical thickness of these clumps is  $ \Delta r = \tau_e / ( \sigma_T
N_e (\mathit{clump}) ) = 0.0005$ and $0.05 R_* $, respectively, at $ r/R_{*} = 1.5 $.  Figure 5 illustrates
this "pancake-like" geometry.  The mass of the thick and thin clumps is the same, a result of the
requirement that they produce the same polarization. The instantaneous stellar mass loss rate due
to a clump compared to that in the base wind is  $ (N(\mathit{clump})/N(\mathit{wind})) \Delta \Omega /4 \pi =
0.16$ and $16 $ at $ r/R_{*} = 1.5 $, which lasts for a time $ \Delta r/v = 1$ and $0.01 $ day,
respectively.  Given that the repetition rate for polarimetric clumps is 20 - 30 days, these clumps
account for at most 1\% of the mass loss of the star. The largest mass clumps are required at the
largest distance from the star: at $ r/R_{*} = 2.5 $, they still account for at most 2\% of the wind
mass loss.
\begin{figure}
\plotfiddle{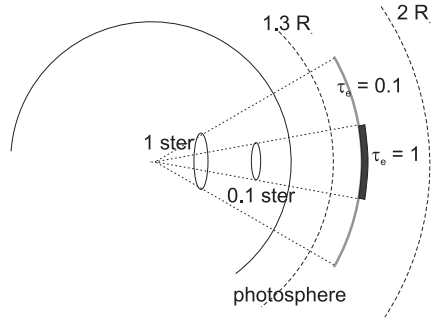}{1.4in}{0}{32}{32}{-115}{-95}
\caption{Clump cartoon, to scale.}
\end{figure}

Finally, we may speculate on the time evolution of the clumps and their polarization.  This will
depend on the evolution of the mass and density of the clumps as they are carried out in the wind. 
The analysis above has placed limits on the properties of the clumps at the time that they
contribute the largest polarization. The polarization will grow as the clump appears out of the
photosphere, possibly growing in mass as it forms.  It may then dissipate and its density, optical
depth, and polarization will drop as it is carried out in the wind.  This is a complex problem
requiring a proper radiative transfer and clump physics model. We may explore the purely
geometric effects, however: if we ignore optical depth effects and changes in clump mass, 
$ p(r) \propto D(r) / r^2 $.  This has a maximum at $ r/R_{\mathit{phot}} \approx 1.1$ $(r/R_* \approx 1.4)
$, and drops by a factor of 10 by $ r/R_* = 3 $.  This is consistent with the polarizing clumps
being highly ionized: by the time they reach the recombined part of the wind, the polarization is
no longer detectable.  The time it takes for a clump to traverse the $ r/R_* = 1.3 - 2.5 $ region
will be $ 1.2 R_* / (55 \mbox{km/s}) = 14$ days, which compares favorably with the observed duration
of roughly one week.

\subsection{Intrinsic Line Polarization}
\begin{figure}
\plotfiddle{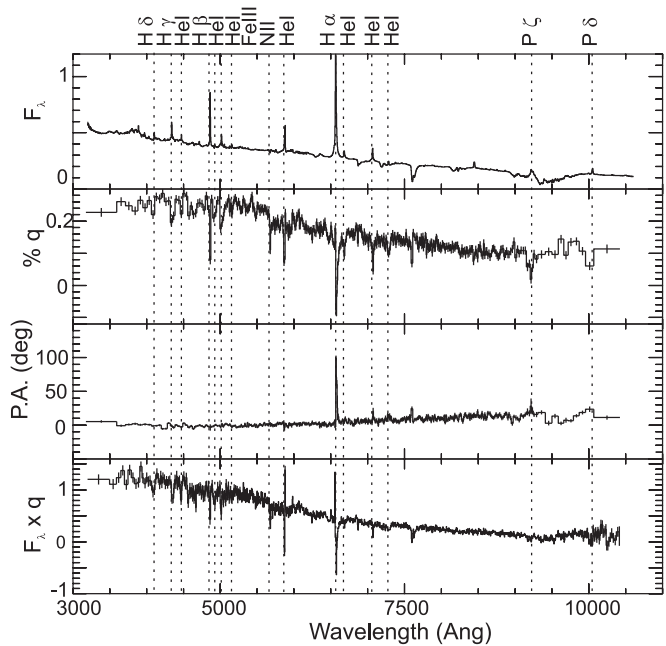}{3.0in}{0}{48}{48}{-150}{-80}
\caption{P Cyg visible wavelength mean polarization wavelength dependence, showing
polarimetric line effects and line identifications.  Top: flux.  Second panel: degree of
polarization.  Third panel: position angle.  Bottom: "polarized flux", showing enhanced
absorption.}
\end{figure}

Figure 6 shows an expanded version of the visible-wavelength mean intrinsic polarization
spectrum from all 15 HPOL observations.   The improved sensitivity of the new HPOL data
allows one to extend the TN analysis to 4 Hydrogen lines and 5 HeI lines.  The intrinsic
polarization of the emission line flux itself is a diagnostic of the wind.  Any line radiation
produced deep in the wind may become polarized if it is scattered off the clumps that produce the
continuum polarization. Alternatively, the P-Cygni absorption beneath the emission line
(unresolved in these observations) may be different in the clump, producing an apparent line
polarization.  The latter appears to be the largest effect in P Cyg.

The line Stokes polarization   $q_{\ell}$  is conventionally computed as follows:
\begin{displaymath}
q_{\ell} = \int ( I( \lambda ) q( \lambda ) - I_c \, q_c )\, d \lambda \, /  \int ( I(\lambda) - I_c ) \, d
\lambda
\end{displaymath}
where the integral is performed over the observed line profile, and $I_c$ and $q_c$ are the flux
and Stokes polarization in the neighboring continuum.  The initially surprising result is that the
weaker lines all appear to be negatively polarized (polarized orthogonal to the continuum).   We
believe that this is an artifact of the fact that the P Cygni absorption is enhanced in the clumps,
and is as a result interpreted as emission polarized orthogonal to the continuum.  For such a
situation one can show that the apparent net Stokes polarization $q_{\ell}$  of the line is
\begin{displaymath}
q_{\ell} =  \frac{q_e - f q_c \, W_a/W_e } { 1 - W_a /W_e },
\end{displaymath}
where  $q_e$ and $q_c$ are the Stokes polarization of the emission line and continuum,
respectively, $W_a$ and $W_e$ are the equivalent widths of the absorption and emission part of
the P-Cygni line, respectively, and  $ f  = W_a (pol) / W_a $ is the factor by which the absorption
line is enhanced in the polarized light.  $W_a$ and $W_e$ were evaluated using the mean P Cyg
line profiles in Stahl, et al (1993).
\begin{figure}
\plotfiddle{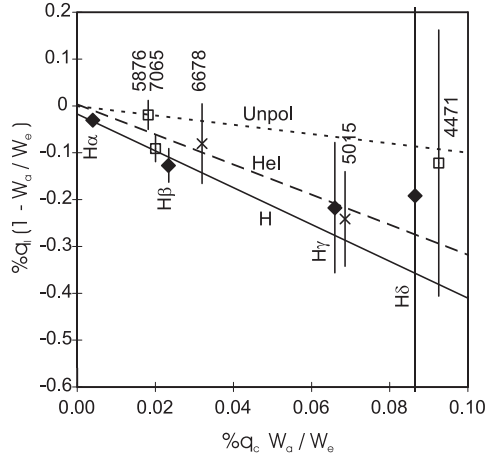}{2.2in}{0}{33}{33}{-120}{-65}
\caption{Line polarization  diagnostic diagram.  Solid diamonds: H;  Crosses HeI (ortho);
squares: HeI (para).  Dashed line: locus for unpolarized emission and unenhanced absorption. 
Solid and dashed lines: unpolarized emission plus enhanced absorption for H and HeI.}
\end{figure}
 
In Figure 7 we plot $q_{\ell} \, (1 - W_a /W_e )$ versus $ q_c \, W_a / W_e $ for the 9 lines
measured.  In this plot, the intercept of a straight line fit to the lines of one species  is  $q_e$  and
the slope is  $f$.  The line labeled "unpol" in this diagram corresponds to $q_e = 0$ and $f = 1$. 
We see that both the H and HeI  lines are consistent with unpolarized emission ($q_e = 0 \pm
0.02\%$), implying that 
{\it more than $90\%$ of the emission arises outside of the polarizing region. }
On the other hand, 
{\it both H and HeI are strongly enhanced in absorption in the polarized flux ($f \sim 4$).}
  The enhancement must be due to a complex combination of ionization and excitation effects,
and will require a detailed analysis.  It is perhaps surprising that the values are so similar for H
and HeI.  Another way of seeing these effects is in the bottom panel of figure 6, the "polarized
flux".  This is essentially the spectrum of the scattered light: the emission lines have almost
vanished, indicating the emission is unscattered, and strong absorption is seen instead.

\section{Relation of Polarimetry to other Observations}

A frequently mentioned datum for P Cyg is the apparent rotation from $ v \sin i $ measurements. 
Markova (2000) favors a value of 40 km/s, and it is agreed that values less than 25 km/sec are
not consistent with absorption line depths.  40 km/s is $15\%$ of breakup velocity for a $30
M_{\sun}$, $75 R_{\sun}$ star (270 km/s).  This is large enough that one might expect a
substantially flattened wind if viewed edge-on.  However, the intrinsic polarization does not
show a favored position angle:  In figure 2a the continuum polarization points occupy a nearly
circular region.  By contrast, Be stars, which are rotating at $\sim50\%$ break-up, show
polarization vector diagrams confined to a narrow line.  One way out of this apparent
contradiction is if P Cyg is rotating rapidly, and is viewed not quite pole-on.  For such an object,
assuming that all clumps are ejected at random in the equatorial plane, the expected axial ratio in
the polarization vector diagram is  $\sim \cos^2 i$.  If we insist that $\cos^2 i > 0.9$, this gives $i
< 18 \deg$.  This in turn gives  $v > v \sin i / 0.3  = 120$ km/s, which is still $45\%$ breakup. 
This would likely be a bipolar object, but viewed nearly pole-on.  Several things recommend this
scenario: First, P Cyg has an unusually high wind speed among S Dor variables (Luminous Blue
Variables).  This would be explained if we are just observing a fast polar wind in a bipolar
object.  Second, most S Dor's are bipolar.  And third, most clumps would be ejected near the
plane of the sky, which is favorable for large polarimetric variations.  On the other hand,
spectroscopy of the P Cyg shell (Meaburn 2000) does not show the characteristic multiple-
redshift components seen in bipolar Planetary Nebulae, for instance.  The pole-on hypothesis is
testable by looking for a systematic daily change in the polarimetric position angle. During the 14
days of detectability, a clump would rotate about 60\deg about the star, and each successive
clump would exhibit the same sense of rotation.

How might the polarimetric clumps be related to the Discrete Absorption Components (DAC's)
seen in P Cyg?  If there were no rotation, or the object is viewed pole-on, they would probably be
uncorrelated, since polarimetric clumps must be ejected near the plane of the sky and would
never cover the stellar disk to form DAC's.  On the other hand, if the star is seen equator-on and
is rotating at $v_{rot} = 40$ km/s, a clump will have rotated about 20\deg by the time it reaches
$2 R_*$, and $60 \deg$ by $10 R_*$ (if the clump maintains its integrity this long).  This
suggests that there would be a weeks - month time delay between the appearance of a
polarimetric clump and its observability as a DAC.  This correlation would vary with the initial
ejection longitude of the clump. Given the relatively frequent generation of polarimetric clumps
(20-30 days), such a variable delay might be difficult to establish.

On the other hand, the polarimetric clumps should certainly be related to $H\alpha$ clumps
observed interferometrically (Chesneau, et al 2000).  If the clump does not disintegrate first, there
should be a roughly 2-6 week delay between the appearance of a polarimetric clump and its
detectability as an $H\alpha$ clump at $10 R_*$.  The position angles of the polarimetric and
interferometric clumps should be correlated, and systematic differences in position angle would
be a sensitive measure of the component of stellar rotation about the line of sight (which
combined with  $v \sin i$ would yield the inclination).

\section{Summary; Future}

In summary, we have analyzed the variability of the wavelength dependence of the polarization
of P Cyg from $1500 - 10500 \AA\ $.  After removal of the interstellar polarization, we have put
limits on the properties of the wind inhomogeneities ("clumps") during the brief period that they
are detectable in polarization.
\begin{itemize}
\item The presence of free-free absorption in the polarization wavelength dependence implies a
clump electron density of at least $10^{13} \mbox{cm}^{-3}$.  This corresponds to a density
contrast of at least 20 at the base of the wind.
\item The absence of Balmer and Paschen bound-free absorption in the polarization limits the
polarimetric zone to $1.3 - 2.5 R_*$, where the hydrogen ionization is high, and requires the
clumps to be substantially cooler than the wind.  The clumps take 14 days to be swept through
this region.
\item In order to produce the maximum observed polarization, the largest clumps must have an
electron optical depth of 0.1 - 1 and an angular extent of 0.1 - 1 ster, accounting for at most
$2\%$ of mean mass loss of the wind.
\item The emission lines are essentially unpolarized.  However, the P Cygni absorption is
enhanced by a factor of four in the polarimetric clumps.
\end{itemize}

We plan on the following further observational investigations:
\begin{itemize}
\item Higher spectral resolution $R \sim 3000$ spectropolarimetry to resolve polarimetric line
profiles.  This should verify the line analysis model of section 3.5.
\item Monitor the polarization every night for at least a month.  During such a campaign we
would look for a change in polarization wavelength dependence during an episode, including the
expected decease in the free-free absorption and the possible appearance of bound-free jumps
near the beginning or end of the episode.  We would also look for a systematic position angle
change with time, indicating rotation.   We will attempt to arrange simultaneous high-resolution
spectroscopy (for DACs) and interferometry, to pursue the predictions of section 4.
\end{itemize}

For modeling, we plan to develop our Monte Carlo polarization radiation transfer code to give a
quantitative continuum polarization model based on the mean wind from Najarro (2000),
including electron scattering, H bound free, H free-free, and line blanketing, with a simple sector
geometry.  This would predict the absolute degree of polarization, the continuum wavelength
dependence, and their time dependence as the clump is carried out in the wind.  Variables would
include the clump heating/ cooling and confinement mechanisms.

A more ambitious polarization model would treat the lines.  To accomplish this, it will be
necessary to more correctly model the ionization and populations in the clump.  Such a model
would predict equivalent width in polarized flux for H and HeI to compare with the observations. 
Finally, our ultimate goal is to calculate polarimetric line profiles, allowing for electron
scattering broadening and Doppler shifts in the wind, to compare with future high resolution
observations.

\acknowledgements

WUPPE was supported by NASA contract NAS5-26777.  The analysis reported in this paper is
supported by a NASA Astrophysics Data Program grant, NAG5-7931.  We gratefully
acknowledge Francisco Najarro for sending early results from his P Cyg wind model, and very
useful discussions with Henny Lamers, Joe Cassinelli, and John Mathis.

\end{document}